# Comments on the question of the discovery of element 112 as early as 1971[*]

R. Brandt

Institut für Physikalische Chemie, Kernchemie und Makromolekulare Chemie
Fachbereich Chemie, Philipps Universität, D-35032 Marburg

## Abstract

There are two independent claims for the discovery of element 112: The claim by Hofmann et al. from 1996 and the older claim from 1971 by Marinov et al. This Comment will not challenge the experimental results of Hofmann et al., but it will discuss one aspect of the claimed discovery of element 112 by Marinov et al., as their experiment has never been reproduced in exactly the form in which the original experiment has been carried out. The reasons for this deficiency may not be found in the field of science, but possibly in radioprotection restrictions for researchers who want to carry out such an experiment. However, such is not a sufficient reason to exclude the original claim from all considerations of the responsible international authorities, who have to settle such priority questions. It may be in agreement with scientific traditions, that when the responsible international committees do not feel to be able to come to a positive decision on the „1971" claim, they could keep the priority problem unsettled for the time being.

-----------

S. Hofmann, et al. published in 1996 a paper (1) on „The new element 112". The reason for this Comment is not to challenge the experimental results of Hofmann et al. but rather to raise the question: „Has the earlier claim of Marinov et al. from 1971 (2,3) received the necessary considerations within the scientific community and the responsible international committees?".

In 1971 evidence (2,3) for the production of element 112 via secondary reactions in CERN W targets was obtained by Marinov et al. The evidence was mainly based on the observation of fission fragments in the Hg sources separated from W targets, on the measured masses of the fissioning nuclei (4) and on the assumption that element 112 (eka-Hg) actually behaves like Hg in the separation process. This problem has been discussed quite extensively over the last decades as being documented also in the recently published literature (5 - 8) and references given therein.

Nobody has been reproducing the original experiment with all essential details. But only an exact reproduction could verify or falsify the original claim with certainty. Many aspects in this entire discovery procedure have been quite specific and original. This applies in particular to the production method via secondary reactions with 24 GeV protons onto a *thick* W target irradiated in CERN. Such an exact repetition of the original experiment has technical and radioprotection problems: A 6 cm long and 30 g heavy W target ( 99.95% purity ) was irradiated for several weeks with a high intensity proton beam of 24 GeV energy at the PS-accelerator in CERN (Geneva) yielding a total fluence of $1*10^{18}$ particles. Such a target is extremely radioactive, nevertheless, the authors managed to start their chemical analytical procedures as early as a few days after the end-of-bombardment. The counting of the alpha-active transuranium samples as well as the chemically isolated ( Hg + 112 ) probe could start about two weeks after the end-of-bombardment. The authors reported the following essential results:

- They started observing about 1 fission event-per-day in the ( Hg + 112 ) sample. This is quite a high activity-rate for these types of experiments. This spontaneous fission activity decayed with an estimated half-life of a few weeks. The authors estimated that the original W-sample contained about 500 atoms of element 112 at the end-of-bombardment - again a remarkable number of new-element atoms in any original discovery for transfermium elements.
- They observed in the chemically separated samples for individual transuranium elements from americium up to lawrencium clear evidences of unknown $\alpha$-activities with many different half-lives ranging up to several years.( 8 - 10)

---

[*] To be published in Kerntechnik.

An early experiment to study these phenomena was carried out by Batty, et al. (11). They used a W target, however, they performed entirely different chemistry procedures ( very specific solvent extraction instead of not so specific ion exchange) and entirely different source preparation procedures ( during their electrolysis the volatile eka-Hg could have easily evaporated ). They observed no spontaneous fission activity ( zero events). But they concluded (quote): „The origin of the fission events observed in the early stages of the W2 experiment (of Marinov, et al. [note added by RB.]) remains unexplained." In particular, it must be stated that Marinov et al. have observed fission fragments several times in two W targets, (W2) and (W3). They observed masses 308 in two different exposures and mass 318 in four different exposures (4).

The closest approach in the exact reproduction of the original „Marinov" experiment has been carried out by Ross, et al. (12). They investigated small W-targets with masses 3.4 up to 5.4 g irradiated with $1.7*10^{17}$ protons at 24 GeV. Therefore, the (fluence*weight) integral was only between 2% up to 3% as compared to the original work. In addition, Ross et al. had to wait 0.3 years up to 1.4 years before they were able to start their chemical processes - this long delay time may have been necessary in order to fulfill local radioprotection requirements. Ross et al. reported the following essential results.

- They observed NO fission fragment activity. The reasons may have been twofold: Based on „Marinov´s" decay rate of 1 fission/day, they could have observed at the best only (0.02 - 0.03) fissions/day. These figures are obtained without taking into account the decay of the activity. Assuming a half-life of one month these figures should be reduced by another factor of at least 16. The background of the fission proportional counters was approximately 0.3 fissions/day and thus quite inadequate to see any possible positive evidence ( the author of this Comment carried out these measurements personally ).

- They observed in the actinide sample definite evidence for unknown α-activities in the energy range of ( 5.43 - 5.49 ) MeV, just as the claims of the „Marinov-Papers" described in (8 - 10). In addition, they observed well-known α-activities from polonium up to thorium. The origin of these known activities could not be given with sufficient certainty: It could be either due to build-up reactions from the W target itself or due to spallation from U/Th impurities within the target. The exact amount of this U/Th-impurity was not known to the authors.

Two further irradiations, this time using massive uranium targets irradiated with 24 GeV protons at the CERN-PS accelerator - similar to the W-irradiation - have been reported in the literature (13,14). Both experiments gave completely negative results with respect to the observation of any new spontaneous fission activity. The investigations concentrated on the chemically evaporated Hg-fraction, as well as on the acid sulfide precipitation in aqueous solution presumably carrying mercury and eka-mercury (element 112). As an example, Ref. 14 describes the investigation of a 65 g metallic uranium target irradiated with $( 4 \pm 1 )*10^{17}$ protons. The chemical process ( dissolution and HgS precipitation ) started 3 weeks after the end-of-bombardment, the HgS precipitate was placed as a thin layer on a mica fission fragment detector for times up to 2.6 years. Afterwards, the HgS was dissolved off the mica and fission fragment tracks were etched chemically. The number of fission fragment tracks was counted using an optical microscope and only ( 1 ± 1 ) track was observed on the entire mica surface. This was just the expected background rate. However, these experiments worked on the assumption that the „Marinov-112" nuclides could also be produced in uranium just as in a tungsten target. It has been shown (4) that the production of the fission activity seen by Marinov et al. (2 - 4) is much less probable in U than in W. A consistent interpretation of the results with the W targets, in terms of the newly discovered super- and hyperdeformed isomeric states are given in Ref. (8). As this Comment concentrates on purely experimental evidences one should mention, that some recent experiments cast some doubts on our complete understanding of nuclear reactions induced by secondary particles within *thick* targets. Such targets could range from copper up to uranium in mass. They could be irradiated with protons or heavier ions with energies above about 10 GeV (15,16).

Recently, some experimental results on the chemical properties of element 112 have been reported (17,18): the properties of element 112 have been measured in the elemental form ( no oxidation, nor any reduction! ). This is in contrast to Marinov et al. who isolated eka-Hg out of an aqueous phase and therefore necessarily in an oxidized state. Again, there is also in this instance no direct comparison with the original 1971 claim possible.

Under these conditions it appears to be within scientific traditions that when the responsible international committees do not feel to be able to come to a positive decision on the „1971" claim, they may consider to keep the settlement of the priority problem unsettled for the time being.

Friendly discussions with Professor A. Marinov in Jerusalem ( Israel ) are kindly acknowledged.